\documentclass[fleqn,twoside]{article}

\usepackage[headings]{espcrc2}

\readRCS
$Id: espcrc2.tex,v 1.2 2004/02/24 11:22:11 spepping Exp $
\usepackage{graphicx}
\usepackage[figuresright]{rotating}

\newcommand{\qq}{Q'} 
\newcommand{\re}{\mathrm{Re}\,}

\newcommand{\gev}{\,{\rm GeV}}

\newcommand{\AmS}{{\protect\the\textfont2
  A\kern-.1667em\lower.5ex\hbox{M}\kern-.125emS}}

\title{Exclusive photoproduction of lepton pairs 
at LHC}

\author{
B. Pire \address{ CPHT, {\'E}cole Polytechnique, CNRS, 91128 Palaiseau, France},
       L. Szymanowski 
        and
J. Wagner \address{ Soltan Institute for Nuclear Studies, Ho\.{z}a 69, 00-681
Warsaw, Poland}}

\begin{document}

\begin{abstract}
Exclusive photoproduction of dileptons, $\gamma p\to
\ell^+\!\ell^- \,p$, will be measured in ultraperipheral collisions at LHC. The mechanism where 
the lepton pair comes from a  heavy timelike photon radiated from a quark
interferes with the pure QED process $\gamma \gamma \to
\ell^+\!\ell^- $.
As an  analog of deeply virtual
Compton scattering, this {\em timelike Compton scattering}  is a way to study generalized parton distributions in the
nucleon or the nucleus.  High energy kinematics will enable to focus on gluon distributions.
Nuclear effects may be scrutinized in heavy ion collisions. 
\end{abstract}

\maketitle

\section{Introduction}

A considerable amount of theoretical and experimental work has recently
been devoted to the study of deeply virtual Compton scattering (DVCS),
 i.e., $\gamma^* p \to \gamma p$, 
an exclusive reaction where generalized parton
distributions (GPDs) factorize from perturbatively calculable coefficient functions, when
the virtuality of the incoming photon is high enough~\cite{Muller:1994fv}.
It is now recognized that the measurement of GPDs should contribute in a decisive way to
our understanding of how quarks and gluons assemble themselves to
hadrons~\cite{gpdrev}. In particular the transverse
location of quarks and gluons become experimentally measurable via the transverse momentum dependence of the GPDs \cite{Burk}.

 The ``inverse'' process, $\gamma p \to \gamma^* p$ at small $t$ and large
\emph{timelike} virtuality $Q'^2$ of the final state photon, timelike Compton scattering 
(TCS) \cite{TCS},
shares many features of DVCS. The Bjorken variable in that case is $\tau = Q'^2/s $.
The possibility to use high energy hadron colliders 
as powerful sources of quasi real photons~\cite{UPC} leads to the hope of determining 
gluonic GPDs in the small skewedness region, which is an essential program complementary  to the determination
 of the quark  GPDs  at lower energy electron accelerators. Moreover, the crossing from 
 a spacelike to a timelike probe is an important test of the understanding of QCD 
 corrections, as shown by the history of the understanding of the Drell-Yan reaction
  in terms of QCD.

\begin{figure}[htb]
\begin{center}
\includegraphics[width=15pc]{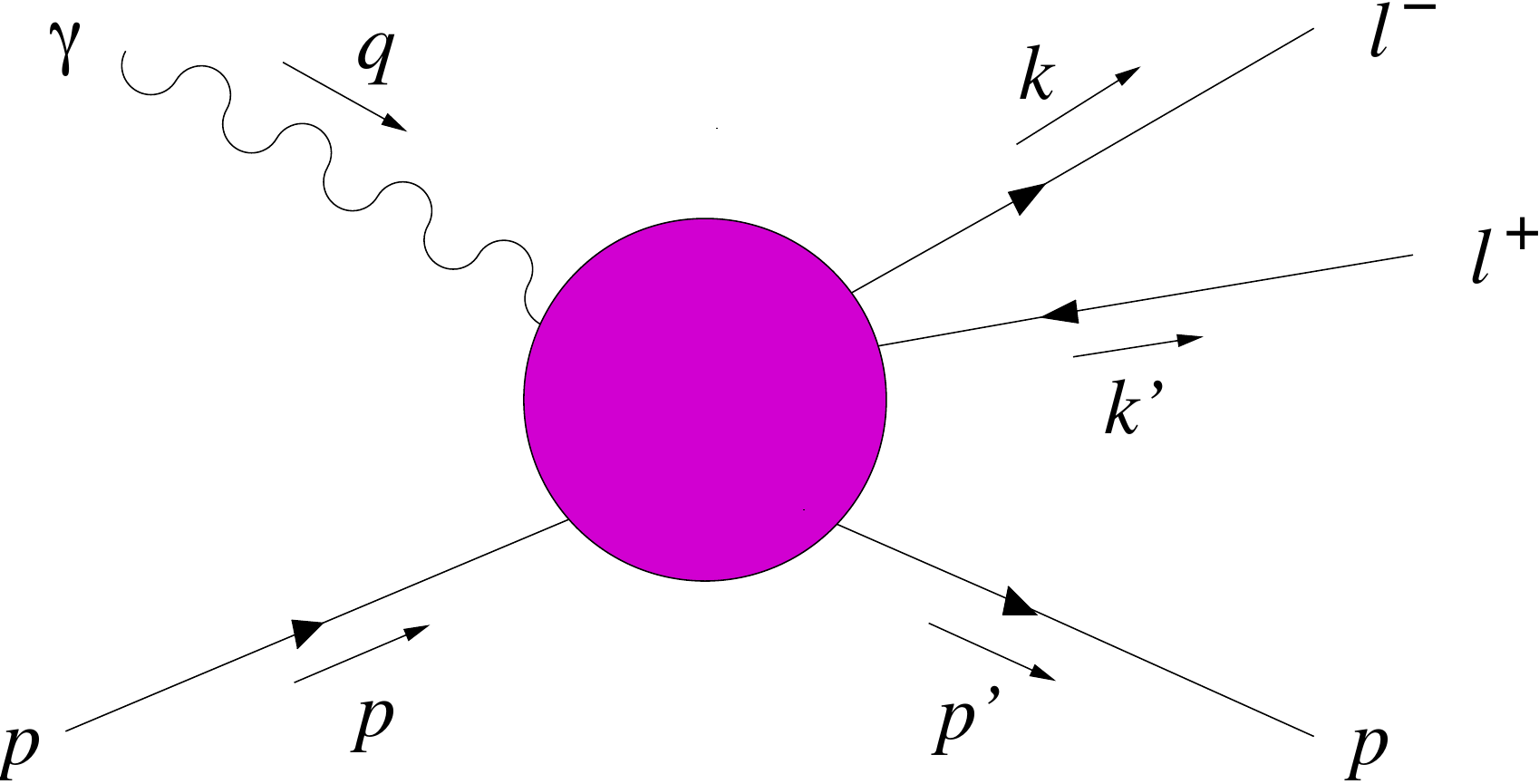}
\caption{\label{refig}Real photon-proton scattering into a lepton pair
and a proton.}
\end{center}
\end{figure}
%

\begin{figure}[tb]
\begin{center}
\includegraphics[width=30pc]{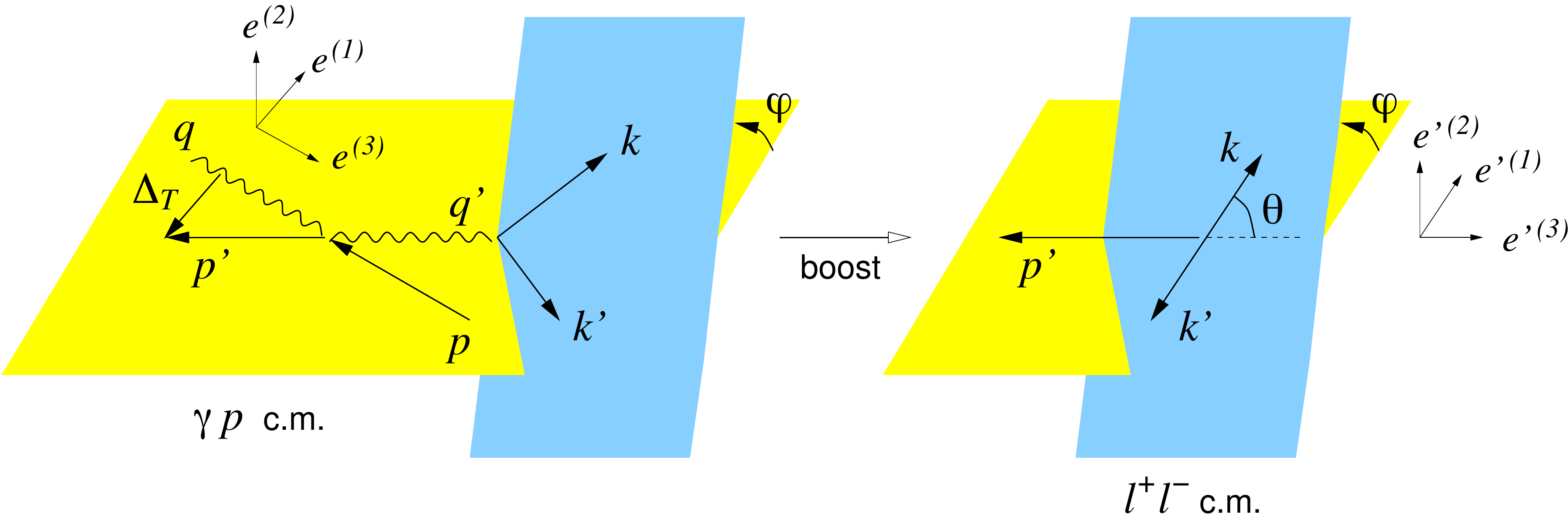}
\caption{Kinematical variables and coordinate axes in
the $\gamma p$ and $\ell^+\ell^-$ c.m.\ frames. }
\label{angle}
\end{center}
\end{figure}

The physical process where to observe TCS is photoproduction of a
heavy lepton pair, $\gamma p \to \mu^+\!\mu^-\, p$ or $\gamma p \to
e^+\!e^-\, p$, shown in Fig.~\ref{refig}. As in the case of DVCS, a Bethe-Heitler (BH)
mechanism contributes at the amplitude level. The interference between the TCS and BH 
processes can readily be accessed through the angular distribution of the lepton pair.
In the $\ell^+\ell^-$ c.m.,  one introduces the polar and azimuthal angles $\theta$
and $\varphi$ of $\vec{k}$, with reference to a coordinate system with
$3$-axis along $-\vec{p}\,'$ and $1$- and $2$-axes such that $\vec{p}$
lies in the $1$-$3$ plane and has a positive
$1$-component.This is shown in Fig. \ref{angle}.

%

\section{The various contributions}
\subsection{The Bethe-Heitler contribution}
\label{sec:bethe}

\begin{figure}[tb]
\begin{center}
\includegraphics[width=15pc]{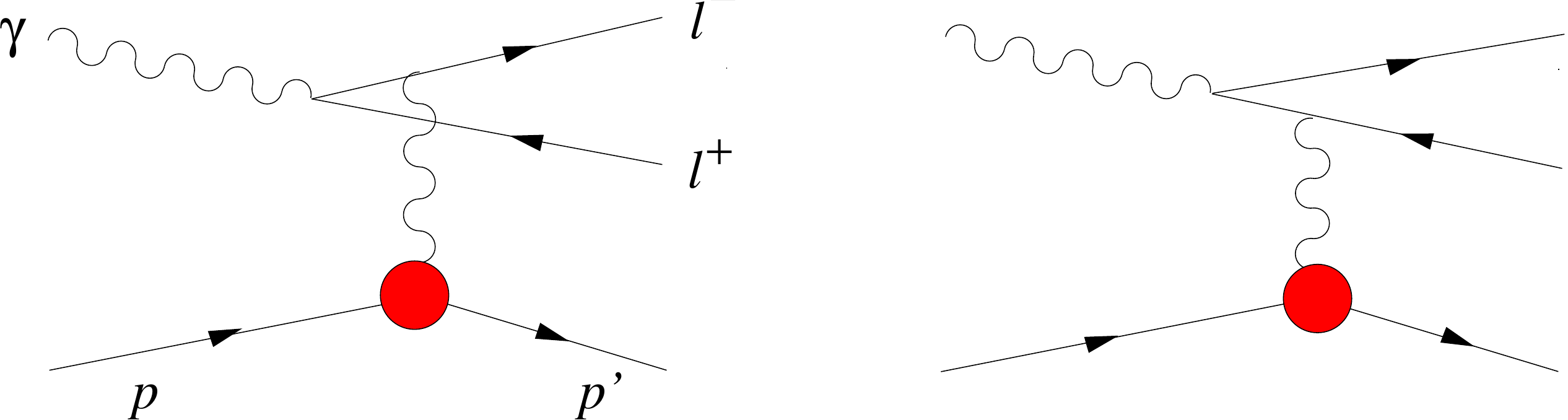}
\caption{The Feynman diagrams for the Bethe-Heitler amplitude.}
\label{bhfig}
\end{center}
\end{figure}

The Bethe-Heitler amplitude is readily calculated from the two Feynman
diagrams in Fig.~\ref{bhfig}. We parameterize the photon-proton vertex
in terms of the usual Dirac and Pauli form factors $F_1(t)$ and
$F_2(t)$, normalizing $F_2(0)$ to be the anomalous magnetic moment of
the target.  Neglecting masses and $t$  compared to
terms going with $s$ or $\qq ^2$, the BH contribution to the unpolarized
$\gamma p$ cross section is ($M$ is the proton mass) 
\begin{eqnarray}
  \label{approx-BH}
&&\frac{d \sigma_{BH}}{d\qq ^2\, dt\, d(\cos\theta)\, d\varphi}
  \approx
    \frac{\alpha^3_{em}}{2\pi s^2}\, \frac{1}{-t}\,
    \frac{1 + \cos^2\theta}{\sin^2\theta} \,
 \nonumber \\
&& \hspace*{-0.3cm}\left[ \Big(F_1^2 -\frac{t}{4M^2} F_2^2\Big)
            \frac{2}{\tau^2}\, \frac{\Delta_T^2}{-t}\,
        + (F_1+F_2)^2 \,\right] ,
\end{eqnarray}
provided we stay away from the kinematical region where the  product  of lepton propagators goes 
\vspace{6.5cm}
.

\noindent
to zero at very small $\theta$.
The interesting physics program thus imposes a
cut on $\theta$ to stay away from the region where the BH cross section becomes
extremely large.

\subsection{The Compton amplitude}
\label{sec:compton}

Both DVCS and TCS are limiting cases of the general Compton process
\begin{equation}
  \gamma^*(q) + p(p) \to \gamma^*(q') + p(p') ,
\label{general-compton}
\end{equation}
where the four-momenta $q$ and $q'$ of the photons can have any
virtuality.  Defining $\Delta = p' - p$, the invariants are
$
Q^2 = - q^2 , \; Q'^2 = q'^2 , \;
s = (p+q)^2 , \; t = \Delta^2 
$
and the scaling variables $\xi$ and $\eta$ read
\begin{eqnarray}
&&\hspace*{-0.3cm}\xi  = - \frac{(q+q')^2}{2(p+p')\cdot (q+q')} \,\approx\,
           \frac{Q^2 - Q'^2}{2s + Q^2 - Q'^2} , \nonumber \\
&&\hspace*{-0.3cm}\eta = - \frac{(q-q')\cdot (q+q')}{(p+p')\cdot (q+q')} \,\approx\,
           \frac{Q^2 + Q'^2}{2s + Q^2 - Q'^2} ,
\label{xi-eta-def}
\end{eqnarray}
where the approximations hold in the kinematical limit we are working
in. $x$, $\xi$, and $\eta$ represent plus-momentum fractions (Light-cone coordinates are defined as $v^\pm = \frac{v^0\pm v^3}{\sqrt{2}}$ , both proton momenta
$p$ and $p'$ moving fast to the right, i.e., having large plus-components).
\begin{equation}
x = \frac{(k+k')^+}{(p+p')^+} , \;
\xi \approx - \frac{(q+q')^+}{(p+p')^+} ,\; 
\eta \approx  \frac{(p-p')^+}{(p+p')^+} .
\nonumber
\end{equation}
To leading-twist accuracy one
has $\xi = \eta$ in DVCS and $\xi = - \eta$ in TCS. 

\begin{figure}
\begin{center}
\includegraphics[width=7pc]{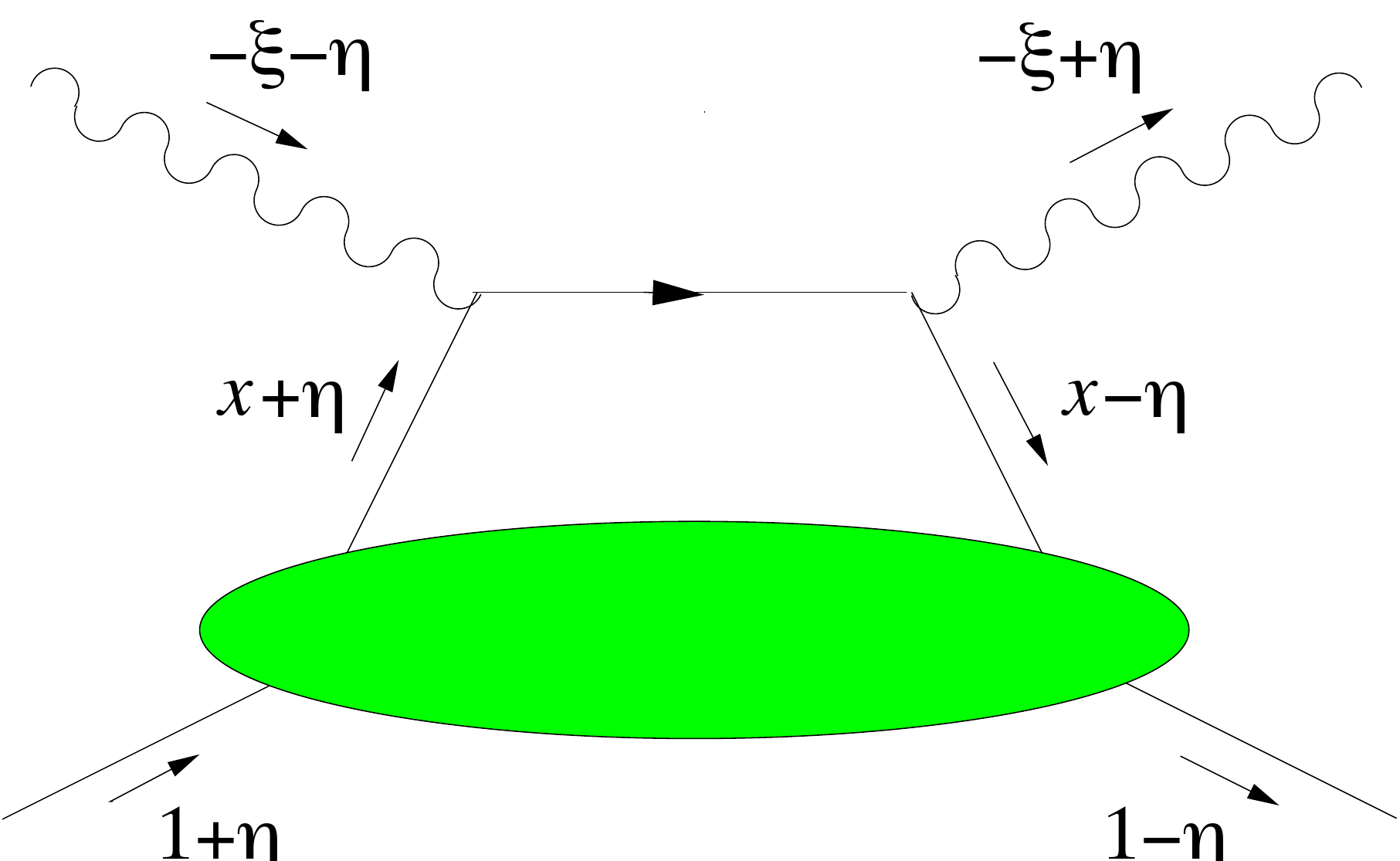}
\hspace{0.05\textwidth}
\includegraphics[width=7pc]{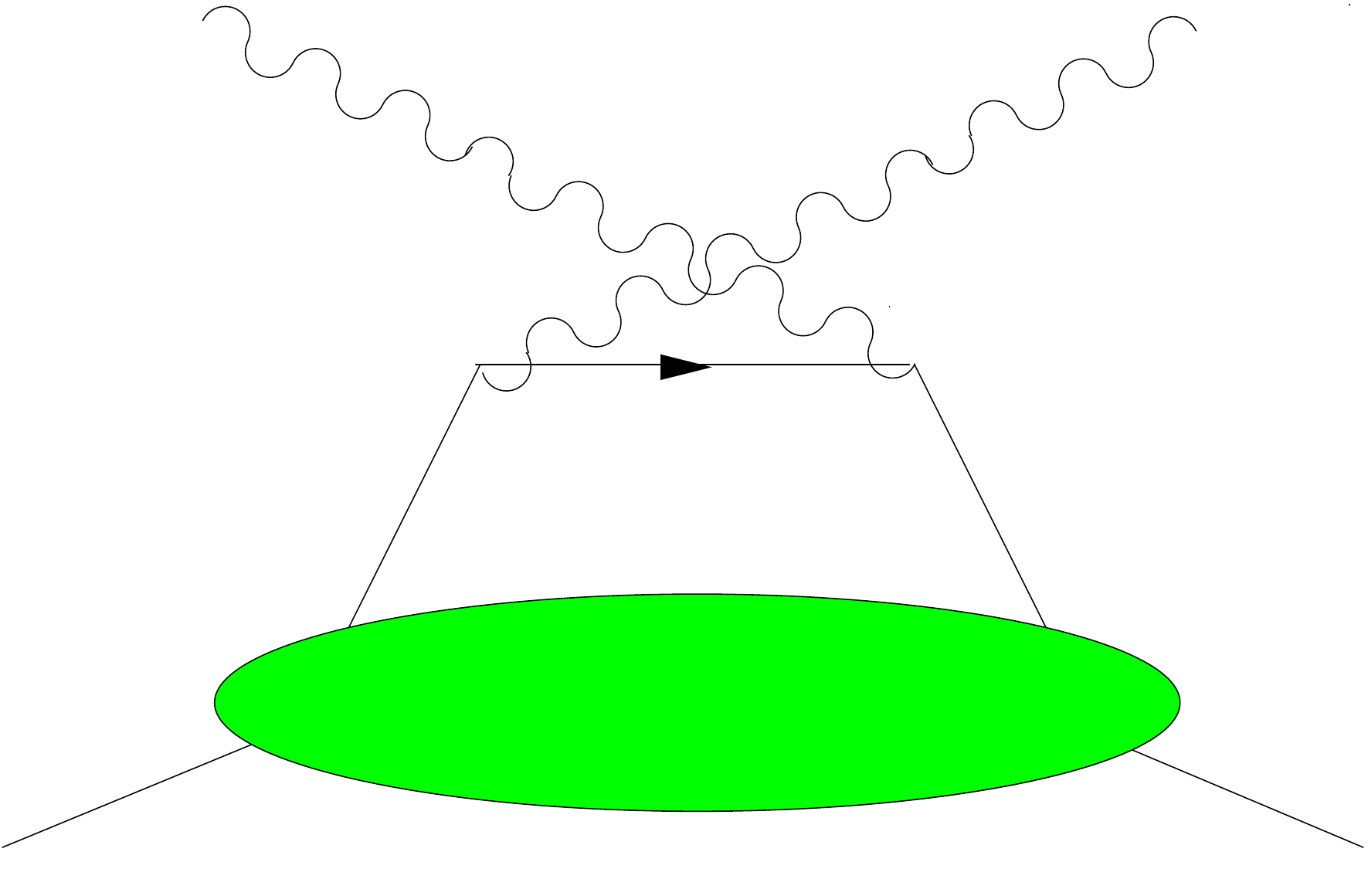}
\caption{Handbag diagrams for the Compton process
(\protect\ref{general-compton}) in the scaling limit. The
plus-momentum fractions $x$, $\xi$, $\eta$ refer to the average proton
momentum $\frac{1}{2}(p+p')$.}
\label{haba}
\end{center}
\end{figure}

 In the region where at least one
of the virtualities is large, the amplitude is given by the
convolution of hard scattering coefficients, calculable in
perturbation theory, and generalized parton distributions, which
describe the nonperturbative physics of the process. To leading order
in $\alpha_s$ one then has the quark handbag diagrams of
Fig.~\ref{haba}.  
The analysis of these handbag diagrams  show the simple relations
\begin{eqnarray}
&& M^{\lambda'+,\lambda+} \Big|_{TCS}
  = \Big[ M^{\lambda'-,\lambda-} \Big]_{DVCS}^* , 
\nonumber \\
&&
M^{\lambda'-,\lambda-} \Big|_{TCS}
  = \Big[ M^{\lambda'+,\lambda+} \Big]_{DVCS}^*
\label{TCS-DVCS-amp}
\end{eqnarray}
between the helicity amplitudes for TCS and DVCS at equal values of
$\eta$ and $t$.  These relations should be evaluated at corresponding
values of $Q'^2$ and $Q^2$ since the photon virtualities play
analogous roles in providing the hard scale of the respective
processes and thus enter in the scale dependence of the parton
distributions.  The relations (\ref{TCS-DVCS-amp}) tell us that at
Born level and to leading twist one obtains the amplitudes for TCS
from those of DVCS by changing the sign of the imaginary part and
reversing the photon polarizations.  To this accuracy, the two
processes thus carry exactly the same information on the generalized
quark distributions. However, the relations (\ref{TCS-DVCS-amp}) no longer hold at $O(\alpha_s)$, neither for the one-loop corrections to the quark handbag diagrams
nor for the diagrams involving gluon distributions. Indeed the TCS amplitude has
discontinuities in both $s$ and $Q'^2$, with one-loop hard scattering
diagrams contributing to both cuts. In situations where $O(\alpha_s)$
contributions are important, the DVCS and TCS processes will have a
different dependence on the generalized parton distributions.  

%

\subsection{The interference term}
\label{sec:inter}

Since the amplitudes for the Compton and Bethe-Heitler
processes transform with opposite signs under reversal of the lepton
charge,  the interference term between TCS and BH is
odd under exchange of the $\ell^+$ and $\ell^-$ momenta, whereas the
individual contributions of the two processes are even. Any observable
that changes sign under $k\leftrightarrow k'$ will hence project out
the interference term, eliminating in particular the eventually large BH
contribution. Clean information on the interference term is therefore
contained in the angular distribution of the lepton pair. 
The interference part of the cross
section for $\gamma p\to \ell^+\ell^-\, p$ with unpolarized protons
and photons is given by

\begin{eqnarray}
  \label{intres}
&&\hspace*{-0.8cm}\frac{d \sigma_{INT}}{d\qq^2\, dt\, d(\cos\theta)\, d\varphi}
= {}-
\frac{\alpha^3_{em}}{4\pi s^2}\, \frac{1}{-t}\, \frac{M}{Q'}\,
\frac{1}{\tau \sqrt{1-\tau}}\,
\nonumber \\
&&\hspace*{-0.8cm}\left[\, \cos\varphi \frac{1+\cos^2\theta}{\sin\theta}
    \re\tilde{M}^{--}
  - \cos2\varphi\, \sqrt{2}\cos\theta\, \re\tilde{M}^{0-}
\right.
\nonumber \\
&&\hspace*{-0.4cm} \left.  + \cos3\varphi\, \sin\theta\, \re\tilde{M}^{+-}
+ O\Big( \frac{1}{Q'} \Big)
\right] ,
\end{eqnarray}
with 
\begin{eqnarray}
&&\tilde{M}^{\mu'\mu} = \frac{\Delta_T}{M}
    \left[ (1-\tau) F_1 - \frac{\tau}{2}\, F_2 \,\right]
       M^{-\mu',-\mu}
\nonumber \\
&&\hspace*{-0.8cm}
+ \frac{\Delta_T}{M} \left[ F_1 + \frac{\tau}{2}\, F_2 \,\right]
       M^{+\mu',+\mu} - \frac{\Delta_T^2}{2M^2}\, F_2\, M^{+\mu',-\mu}
\nonumber \\
&&\hspace*{-0.8cm} {}+ \left[ \tau^2 (F_1+F_2) + \frac{\Delta_T^2}{2M^2} F_2 \,\right]
       M^{-\mu',+\mu}
\end{eqnarray}
is the same combination of Compton helicity amplitudes as defined in
\cite{Diehl:1997bu}. The close analogy between TCS and DVCS is manifest, and  a
$\gamma^*$ with negative helicity in TCS corresponds to a $\gamma^*$
with positive helicity in DVCS as  already found in the relations
(\ref{TCS-DVCS-amp}).
Without polarization, one probes the real parts of the
Compton helicity amplitudes. 

\begin{figure}
\begin{center}
\includegraphics[width=15pc]{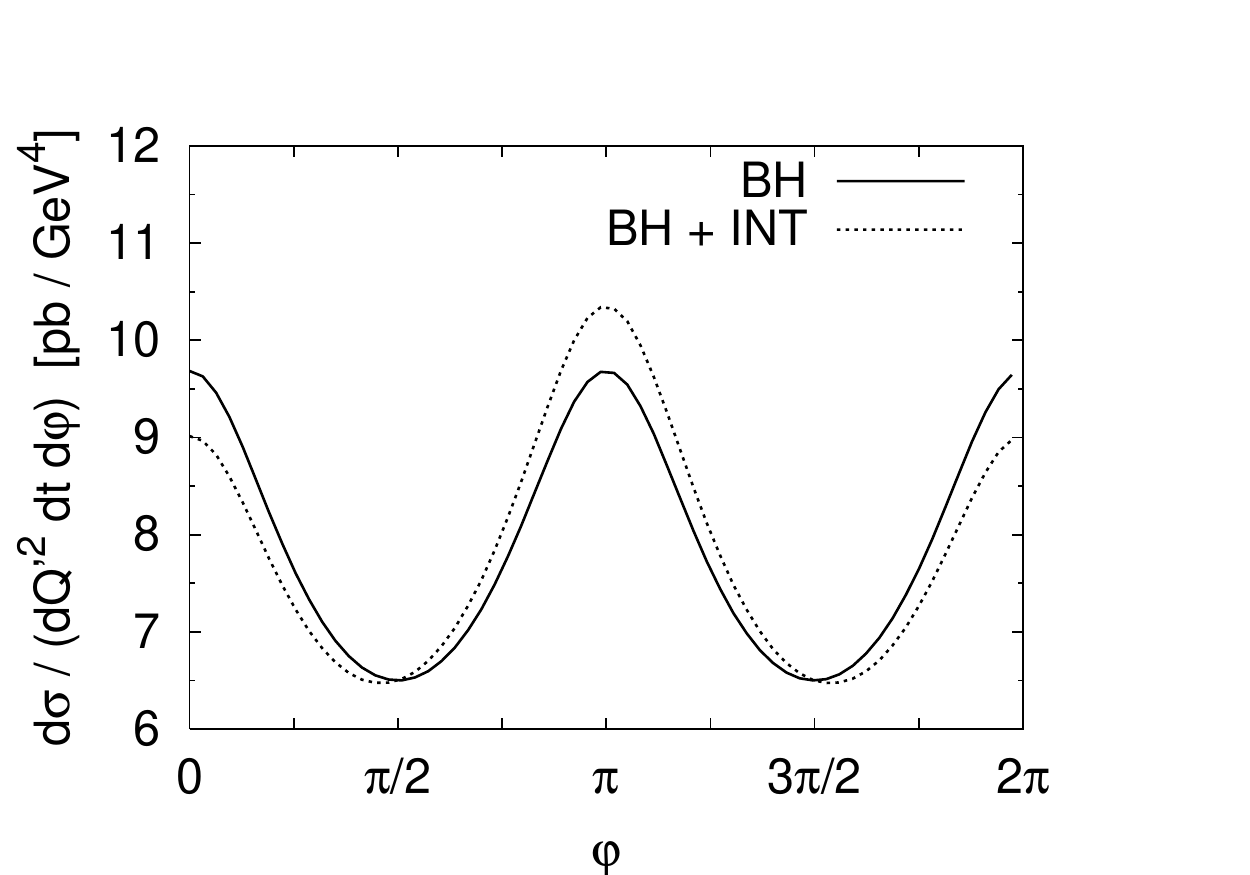}
\caption{ The cross section integrated over $\theta \in [\pi/4,3
\pi/4]$ as a function of $\varphi$ for $\sqrt{s}=5 \gev$, $Q'^2=5
\gev^2$, $|t| =0.2 \gev^2$.  The curves represent the BH contribution
(solid line) and the sum of BH and the interference term (dash-dotted
line).} 
\label{phi}
\end{center}
\end{figure}

Let us summarize the results of numerical estimates obtained in Ref. \cite{TCS}
in the case of low scattering energy. A model calculation  gives the results shown 
on Fig.~\ref{phi}
for the $\varphi$ dependence of the cross
section integrated over $\theta$ in the range $[\pi/4,3 \pi/4]$. With
the integration limits symmetric about $\theta=\pi/2$ the interference
term is odd under $\varphi\to \pi+\varphi$ due to charge conjugation,
whereas the TCS and BH cross sections are even. The
contributions from BH and the sum of BH and the interference term are separately shown.
 The TCS cross section is flat in $\varphi$ to leading-twist accuracy. In the kinematics of 
the figure one gets
$d\sigma_{\it TCS} /(d\qq^2\, dt\, d\varphi) \approx
0.2~\mbox{pb\,GeV}^{-4}$ when applying the same cut in $\theta$.

\begin{figure}
\begin{center}
\includegraphics[width=15pc]{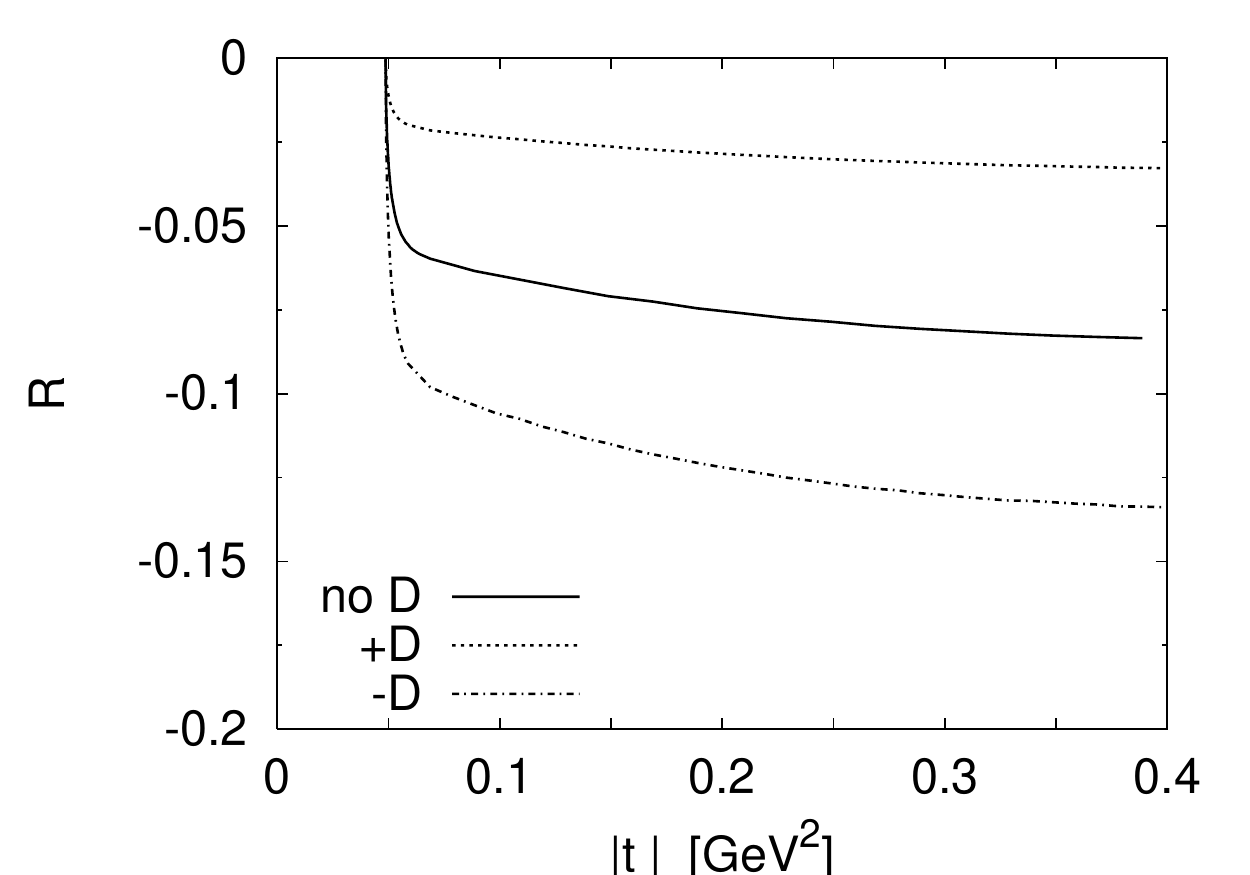}
 \caption{The ratio $R$ defined in (\protect\ref{resnod}).
 for $\sqrt{s}=5 \gev$ and $Q'^2=5 \gev^2$. 
 The curves correspond to three models for the GPDs \cite{TCS}.}
 \label{phiR}
\end{center}
\end{figure}

To extract information on the Compton amplitude in a compact way, 
 Ref.~\cite{TCS} demonstrated that it was useful to introduce the ratio $R$ :
\begin{equation}
\label{resnod}
R = \frac{\displaystyle 2 \int_0^{2\pi} d\varphi\, \cos\varphi\,
          \frac{dS}{d\qq^2\, dt\, d\varphi}}{\displaystyle
\int_0^{2\pi} d\varphi\, \frac{dS}{d\qq^2\, dt\, d\varphi}} \; ,
\end{equation}
which projects out the ratio $a_1 /a_0$ of Fourier coefficients in the
weighted cross section $dS /(d\qq^2\, dt\, d\varphi) =
\sum_{n=0}^{\infty}\, a_n \cos(n\varphi)$.  Up to $1/Q'$ suppressed
contaminations the numerator in $R$ is proportional to the combination
$\tilde{M}^{--}$ of Compton amplitudes, whereas the denominator is
 dominated by the BH part of the cross section. Fig.~\ref{phiR} shows 
this ratio as a function of $t$ for three models of the GPDs.

Concerning  the high energy domain, we are currently working to get an estimate of the 
different contributions to the lepton pair cross section for ultraperipheral collisions at the LHC.  Since the
cross sections decrease rapidly with $Q'^2$, we are interested in the kinematics of moderate $Q'^2$, 
say a few GeV$^2$, and large energy, thus very small values of $\eta$. 
A rough estimate of the Bethe Heitler cross section gives 28 picobarns 
when it is integrated over $\theta$ in the range $[\pi/4,3 \pi/4]$, $\varphi$ in the range $[0, 2 \pi]$,
 $-t $ in the range $[0.05, 0.25]$GeV$^2$ and  $Q'^2 $ in the range $[4.5, 5.5]$ GeV$^2$. This gives hope for a 
 measurable process. It remains to check whether one may detect the modulation of this cross section by the
  interference with the TCS amplitude. The crucial ingredient is realistic models of GPDs at small skewedness.
  Singlet quark and gluon GPDs
will give the dominant contributions to the TCS amplitude in that domain. Since gluon GPDs only enter the TCS 
amplitude at the $O(\alpha_{S})$ level, a consistent treatment requires to take into account GPD evolution equations. 
This raises the question of the choice of factorization scale in a process where the large scale is timelike, which has 
been much debated in the inclusive case of Drell Yan lepton pair production, where enhancement K- factors are usually
understood as coming from the analytical continuation of $log(Q'^2)$ terms from spacelike to timelike values.
Such factors are likely to be present also in our case.

\section{Nuclear targets} 

The operation of LHC as a heavy ion collider will enable us to study TCS on nuclei. Such scattering may be 
coherent and one then needs to define nuclear GPDs \cite{NucGPD}. This is a very interesting subject 
which definitely needs more work. Incoherent TCS will also occur on quasi free neutrons and protons. Let us
 remark  that the BH process is suppressed for a neutron target, due to the zero charge of the neutron. 

\section*{Acknowledgements} 
We acknowledge useful discussions with M. Diehl, D. M\"uller and M. Strikman. This work is partly 
supported by the French-Polish scientific agreement Polonium and by  
the ECO-NET program, contract 12584QK.



\begin{thebibliography}{99}


\bibitem{Muller:1994fv}
D.~M{\"u}ller {\em et al.},
Fortsch.\ Phys.\  {\bf 42}, 101 (1994);
X.~Ji,
Phys.\ Rev.\ Lett.\  {\bf 78}, 610 (1997);
A.~V.~Radyushkin,
Phys.\ Rev.\  {\bf D56}, 5524 (1997);
J.~C.~Collins and A.~Freund,
Phys.\ Rev.\  {\bf D59}, 074009 (1999).

\bibitem{gpdrev}
M.~Diehl,
  Phys.\ Rept.\  {\bf 388} (2003) 41;
  A.~V.~Belitsky and A.~V.~Radyushkin,
  Phys.\ Rept.\  {\bf 418}, 1 (2005);
  S.~Boffi and B.~Pasquini,
  arXiv:0711.2625 [hep-ph].
  
\bibitem{Burk}
M.~Burkardt,
  Phys.\ Rev.\  D {\bf 62}, 071503 (2000)
  and
  Int.\ J.\ Mod.\ Phys.\  A {\bf 18}, 173 (2003);
J.~P.~Ralston and B.~Pire,
  Phys.\ Rev.\  D {\bf 66}, 111501 (2002);
  M.~Diehl,
  Eur.\ Phys.\ J.\  C {\bf 25}, 223 (2002).

\bibitem{TCS}
E.~R.~Berger, M.~Diehl and B.~Pire,
  Eur.\ Phys.\ J.\  C {\bf 23}, 675 (2002).
  
  \bibitem{UPC}
K.~Hencken {\it et al.},
  Phys.\ Rept.\  {\bf 458}, 1 (2008).
  
\bibitem{Diehl:1997bu}
M.~Diehl {\em et al.},
Phys.\ Lett.\  {\bf B411}, 193 (1997).

\bibitem{NucGPD}
E. Berger  {\em et al.},  Phys.\ Rev.\ Lett.\  {\bf 87}, 142302 (2001);
 F.~Cano and B.~Pire,
  Eur.\ Phys.\ J.\  A {\bf 19}, 423 (2004);
  A.~Kirchner and D.~Mueller,
  Eur.\ Phys.\ J.\  C {\bf 32}, 347 (2003);
V.~Guzey and M.~Strikman,
  Phys.\ Rev.\  C {\bf 68}, 015204 (2003);
V.~Guzey,
  arXiv:0801.3235 [nucl-th] and J.\ Phys.\ G {\bf 32}, 251 (2006).
\end{thebibliography}
\end{document}